# Epitaxial InGaAsP/InP photodiode for registration of InP scintillation

**S. Luryi[*], A. Kastalsky, M. Gouzman, N. Lifshitz, O. Semyonov, M. Stanacevic, A. Subashiev, V. Kuzminsky, W. Cheng, V. Smagin, Z. Chen**

*University at Stony Brook, ECE Department and NY State Center for Advanced Sensor Technology, Stony Brook, NY 11794-2350*

**J. H. Abeles, W. K. Chan, Z. A. Shellenbarger**

*Sarnoff Corporation, Princeton, NJ 08540*

**Abstract**

Operation of semiconductor scintillators requires optically-tight integration of the photoreceiver system on the surface of the scintillator slab. We have implemented an efficient and fast quaternary InGaAsP *pin* photodiode, epitaxially grown upon the surface of an InP scintillator wafer and sensitive to InP luminescence. The diode is characterized by an extremely low room-temperature dark current, about 1 nA/cm$^2$ at the reverse bias of 2 V. The low leakage makes possible a sensitive readout circuitry even though the diode has a large area (1 mm × 1 mm) and therefore large capacitance (50 pF). Results of electrical, optical and radiation testing of the diodes are presented. Detection of individual α-particles and γ-photons is demonstrated.

---

[*] Corresponding author. *Email address*: Serge.Luryi@StonyBrook.edu

## 1. Introduction

There are two large groups of solid-state radiation detectors, which dominate the area of ionizing radiation measurements: scintillation detectors and semiconductor diodes. The scintillators detect high-energy radiation through generation of light which is subsequently registered by a photo-detector that converts light into an electrical signal. Semiconductor diodes employ reverse biased p-n junctions where the absorbed radiation creates electrons and holes, which are separated by the junction field thereby producing a direct electrical response. Both groups are extensively reviewed by Knoll [1].

Normally, scintillators are not made of semiconductor material. The key issue in implementing a semiconductor scintillator is how to make the material transmit its own infrared luminescence, so that photons generated deep inside the semiconductor slab could reach its surface without tangible attenuation. However, semiconductors are usually opaque at wavelengths corresponding to their radiative emission spectrum. Our group has been working on the implementation of high-energy radiation detectors based on direct-gap semiconductor scintillator wafers, like InP or GaAs. For the exemplary case of InP the scintillation spectrum is a band of wavelengths near 920 nm. The original idea [2] was to make InP relatively transparent to this radiation by doping it heavily with donor impurities, so as to introduce the Burstein shift between the emission and the absorption spectra. Another approach [3] is based on the extremely high radiative efficiency of high-quality direct-gap semiconductors, such as InP. In these materials, an act of interband absorption does not finish off a scintillation photon; it merely creates a new minority carrier and then a new photon in a random direction. The efficiency of photon collection in direct-gap semiconductors is therefore limited only by parasitic processes, such as nonradiative recombination of the minority carriers and free-carrier absorption of light. If these are minimized, one can have an opaque but "ideal" (in terms of the photon collection efficiency) semiconductor scintillator [3].

Most scintillators reported in the literature are implemented in wide-gap insulating materials doped ("activated") with radiation centers. A classical example of a solid-state scintillator is sodium iodide activated with thallium (NaI:Tl). Because of the much longer wavelength of the scintillation associated with the activator energy levels — compared to the interband absorption threshold — the insulating scintillators are highly transparent to their own luminescence. However, this advantage comes at a price in the transport of carriers to the activator site. Individual carriers have very poor mobility in insulators and transport efficiency requires that the generated electrons and holes form excitons and travel to the radiation site as neutral entities. Therein lies a problem. The energy resolution even in the best modern scintillators does not compare well with that in semiconductors [4]. One of the fundamental reasons for poor resolution is that the luminescent yield in dielectric scintillators is controlled by reactions that are nonlinear in the density of generated electron-hole pairs, such as the formation of excitons at low densities and the Auger recombination at high densities [5-8].

Such nonlinear processes do not exist in doped semiconductors, where interaction with gamma radiation induces *minority* carriers, while the concentration of majority carriers does not measurably change. Every reaction on the way to luminescence, including

Auger recombination, is *linear* with respect to the concentration of minority carriers. One can therefore expect that semiconductor scintillators will not exhibit effects of non-proportionality and their ultimate energy resolution could be on par with that of diode detectors implemented in the same material.

The proportionality of scintillation yield is not the only expected advantage of semiconductor scintillators. One of the major benefits of semiconductor materials is the mature technology that enables the implementation of epitaxial photodiodes integrated on the surface of a semiconductor slab. Such a development is reported in this paper.

We remark that an external receiver, like a photomultiplier, is not a viable option because of the complete internal reflection of most of the scintillating radiation. Owing to the high refractive index of semiconductors, e.g., $n = 3.3$ for InP, most of the scintillating photons will suffer a complete internal reflection at the surface. Only those photons that are incident on the semiconductor-air interface within a narrow cone $\sin \theta < 1/n$ off the perpendicular to the interface, have a chance to escape from the semiconductor. The escape cone accommodates only a small fraction of isotropic scintillation, $\sin^2 \theta/2 < 1/4n^2 \approx 2\%$, whence the inefficiency of collection.

It is therefore imperative to integrate the scintillator wafer with a photodetector that has a substantially similar or even higher refractive index in an optically tight fashion. This paper reports the implementation of *epitaxial photodetectors* on InP scintillator body, implemented as ultra-low leakage *pin* diodes based on quaternary InGaAsP materials. The epitaxial diode provides nearly perfect registration efficiency of photons that have reached the heterointerface.

A semiconductor scintillator endowed with an integrated photoreceiver can be patterned into a two-dimensional array of pixels. Such an array forms a basic unit that can be stacked up indefinitely in the third direction [3]. This in turn enables three-dimensional (3D) integration of scintillator "voxels" (3D pixels). A gamma photon incident on such a 3D array produces a cluster of firing voxels that report their positions and the energy deposited. This report carries valuable information. There is well-developed technique, known as the Compton telescope [9,10], that enables one to deduce both the incident energy $E_0$ and the incident direction cosine $\hat{n}_0$ from the known energies $\Delta_1$ and $\Delta_2$ deposited in the first two voxels and the direction $\theta_2$ to the third voxel.

Semiconductor scintillators offer a tantalizing possibility of implementing a compact low-voltage Compton telescope [3]. One of the keys to this goal is the implementation of optically-tight photoreceiver system on the surface of semiconductor scintillators.

## 2. Design and fabrication

In a 2D array the lateral size of an individual pixel is limited by the increasing capacitance of a photodiode, which limits its sensitivity, as discussed in Sect. 4. For a single-pixel device reported in this work it is desirable that the linear lateral dimensions exceed the thickness of the scintillator, so that the photon collection will be dominated by the planar effects and not by the side-wall reflection. We chose to work with 350 μm thick InP scintillator of 1 mm × 1 mm lateral dimensions.

The resultant large photo-diode area is challenging from the standpoint of dark current of a *pin* diode under reverse bias. Especially troublesome would be the surface leakage at the diode sidewalls, as illustrated in Fig. 1.

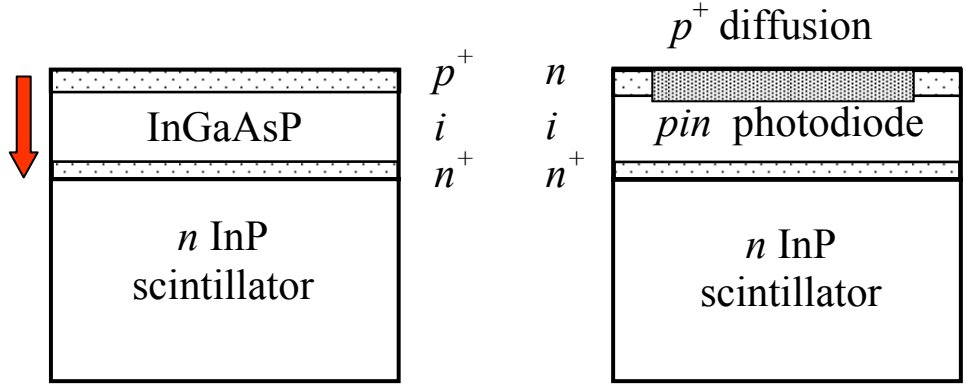

**Fig. 1.** The epitaxial *pin* diode designs. The left panel shows a diode that would have unacceptable leakage, due to a parasitic current path on the diode sidewall (indicated by the arrow). The right panel shows the better design where the sidewall is no longer under bias. The epitaxially grown diode structure is *nin* rather than *pin*, with the $p^+$ region subsequently introduced by diffusion of acceptors (Zn).

Surface leakage is a known problem that had been encountered and overcome years ago in the development of InGaAs *pin* photodetectors and avalanche photodiodes (APD) for optical communications, where the low dark current requirement is paramount, as it is for us. The best solution found by the pioneer researchers [11,12] was to avoid putting the sidewall surface under voltage. This is accomplished by the epitaxial growth of *nin* rather than *pin* structures, with the $p^+$ region subsequently introduced by Zn diffusion only *in the interior* of the diode – some distance $x$ away from the eventual sidewall. In this design one has two *pin* diodes in parallel, one vertical under the diffusion, the other lateral. The distance $x$ to the cleavage plane should be chosen so that the top $n$ layer is undepleted near the sidewall. The latter is therefore in equilibrium at a constant potential and has no current flowing.

The bandgap of undoped InP at 300K is $E_G = 1.35\,\text{eV}$. The photoreceiver bandgap must be lower than that of doped InP, so that the entire luminescence spectrum is absorbed. On the other hand, it should not be narrower than necessary, because for narrower gaps, the photoreceiver noise increases due to the thermal dark current. We have chosen the photoreceiver bandgap $E_G^{Ph} = 1.24\,\text{eV}$, which provides an absorption length of less than 1 μm for InP luminescence.

The active (*i*) region of the *pin* epitaxial photo-diode has been implemented as a quaternary $\text{In}_{1-x}\text{Ga}_x\text{As}_y\text{P}_{1-y}$ alloy lattice-matched to InP. Lattice matching is important because otherwise dislocations will have a detrimental effect on the photoreceiver performance. For lattice-matched compositions ($x = 0.454y$) the alloy bandgap is given by [13] $E_G(y) = 1.35 - 0.72y + 0.12y^2$, so that the desired value is achieved with $x = 0.07$, $y = 0.16$. The epitaxial structure grown by organo-metallic chemical vapor deposition (OMCVD) is illustrated in Fig. 2.

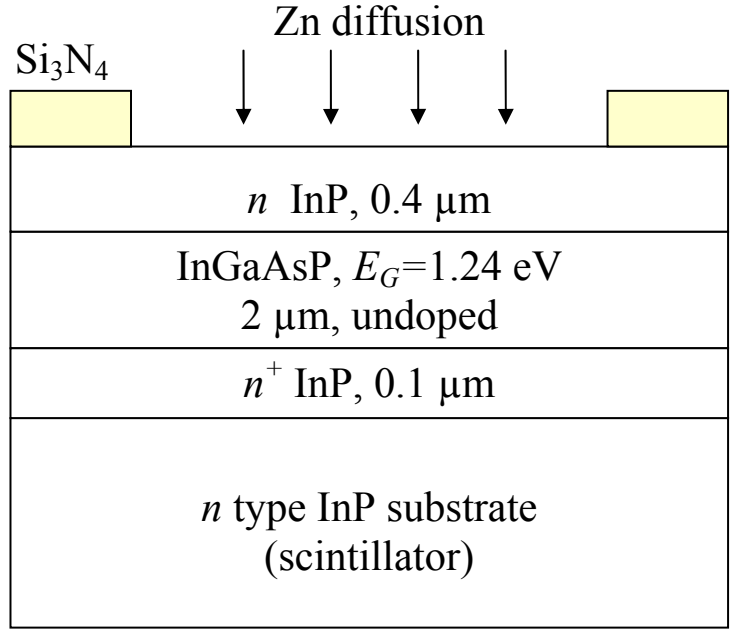

**Fig. 2.** The epitaxial structure cross-section. The active (*i*) layer of the *pin* diode is implemented as a quaternary $In_{1-x}Ga_xAs_yP_{1-y}$ alloy lattice-matched to InP ($x = 0.454y$) with bandgap of 1.24 eV ($x = 0.07$, $y = 0.16$). Thickness (0.4 µm) of the top *n*-type InP layer is chosen so that the diffusion front of Zn penetrates into the quaternary layer, but only slightly. The lateral diode dimension is determined by an opening in the $Si_3N_4$ protective layer and is a 1 mm × 1 mm square with rounded corners.

The diffusion of Zn was carried out in the OMCVD reactor using a silicon nitride mask, as illustrated in Fig. 2. It is important that Zn penetrates all the way through the top *n*-type InP layer and into the quaternary layer, so that the InP/InGaAsP interface is within the $p^+$ doped region. Otherwise, the heterointerface would introduce an unwelcome series resistance. On the other hand, the front edge of the Zn penetration should not be too deep, for that would enhance the diode capacitance and diminish the detector sensitivity. Using secondary ion mass spectrometry (SIMS), we ascertained that 0.5 hour Zn diffusion in InP at 525°C results in the front edge penetration of slightly under 500 nm, so that the appropriate thickness of InP cap layer is 400 nm. The measured SIMS profile, shown in Fig. 3, guides the design of top epitaxial layers of the diode.

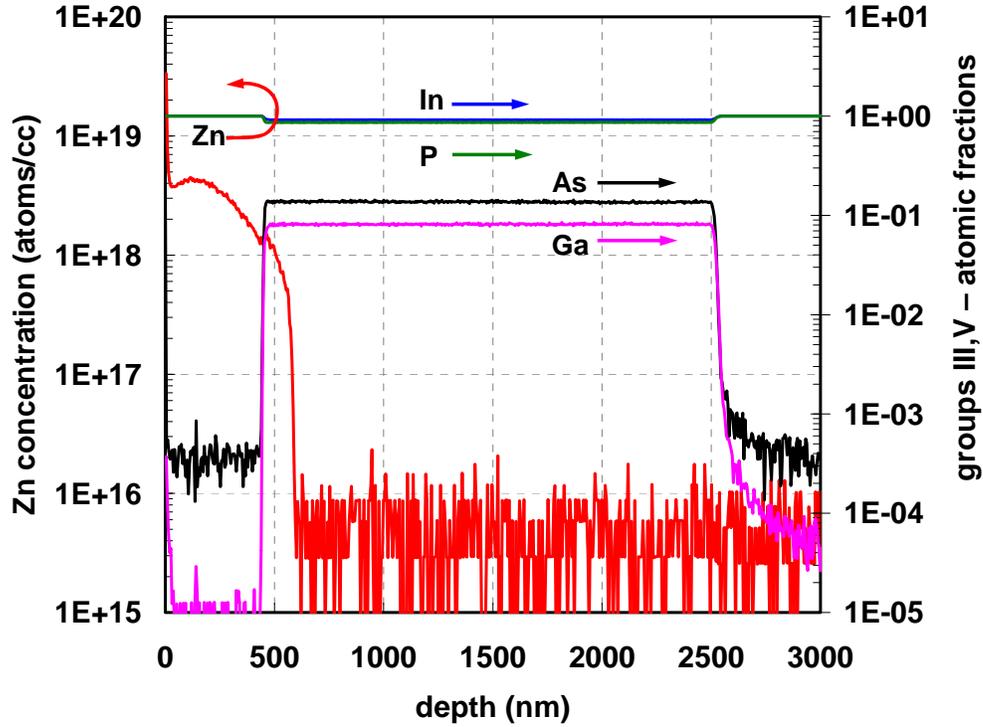

**Fig. 3**. SIMS profile of the heterostructure illustrated in Fig. 2 upon Zn diffusion.

## 3. Photodiode testing

Current-voltage characteristics of the epitaxial diodes are shown in Fig. 4. The reverse-bias current of the diodes is remarkably low, as low as 1 pA at $V_R = -1$V at room temperature. The reverse bias leakage varies from device to device but mostly remains below 10 pA, corresponding to the very low current density of 1 nA/cm$^2$.

The measured voltages for different temperature curves at the same low values of current allows us to estimate the build-in potential, $V_{bi} \approx 1$ V, which is less than the bandgap of the quaternary material (1.25 eV). This estimate is made in a simple thermionic model of barrier emission illustrated in Fig. 5. It is important because it does not assume any *a priori* value for the diode ideality factor and gives us a definite conclusion about the band structure of the quaternary diode.

The argument goes as follows. We make an assumption that the characteristics are exponential, with some "ideality" factor *m*,

$$I \propto \exp\left(\frac{E_A(T) - eV}{2mkT}\right) \qquad (1)$$

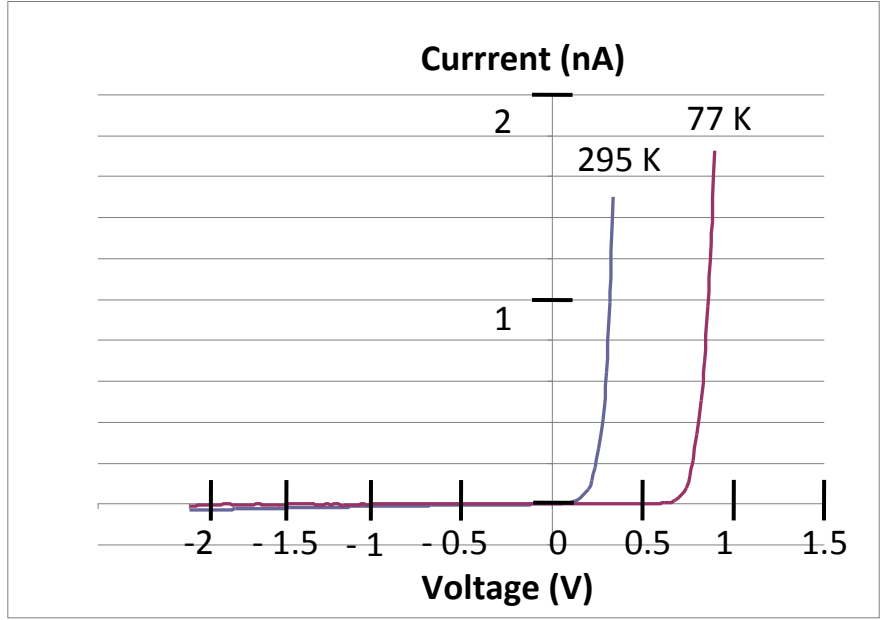

**Fig. 4**. Current-voltage characteristics of the quaternary $Ga_xIn_{1-x}As_yP_{1-y}$ *pin* diodes of area 1 mm × 1 mm. Separation between the turn-on voltages at $T$ = 295 K and 77 K is 0.56 V. Reverse-bias current at room temperature varies between a record low value of 1 pA and about 10 pA in typical diodes at $V_R = -1$ V.

The same value of current is obtained at $V_1(T_1)$ and $V_2(T_2)$ where $V_2 - V_1 = 0.56$ V. The activation energy $E_A$ depends on the temperature $T$ because it additively includes the bandgap. The bandgap difference $\Delta \equiv E_G(T_2) - E_G(T_1) \approx 0.06$ eV for InGaAsP can be estimated from the data for InP:

$$E_G(T) = E_G(0) - \frac{T^2 \times 4.9 \cdot 10^{-4} \text{ eV}}{T + 327} \qquad (2)$$

Equating the currents at some low value in Fig. 4, measuring the forward bias voltages $V_1(T_1)$ and $V_2(T_2)$ and taking the ratio $(T_1/T_2) = 3.83$, we find the experimental vale of the room-temperature activation energy $E_A \approx 1$ eV. We interpret this result by noting that if the nominally intrinsic layer of the *pin* diode is in fact partially undepleted, then the activation energy $E_A$ will differ from the bandgap $E_G$ by the Fermi energy $E_F$ in the unintentionally doped intrinsic layer, see Fig. 5.

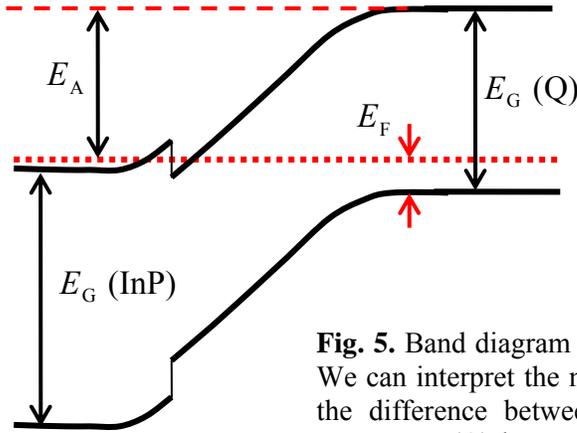

**Fig. 5.** Band diagram of the nominally *pin* layer in equilibrium. We can interpret the measured activation energy, $E_A \approx 1$ eV, as the difference between the bandgap $E_G = 1.24$ eV of the quaternary (Q) layer and the Fermi energy $E_F = 0.23$ eV in the undepleted portion of the quaternary layer.

The band diagram in Fig. 5 is consistent with the impurity doping profile measured by SIMS (Fig. 3) and is also ascertained by the capacitance-voltage (CV) characteristics. The measured capacitance of 63 pF corresponds to an undepleted layer of 1.75 μm at zero bias with the impurity concentration of $10^{15}$ cm$^{-3}$ at the depletion edge. With applied reverse bias, the capacitance decreases to 56 pF as the depletion width widens to about 1.9 μm at 1.5 V. The dopant concentration at the depletion edge increases to $3 \times 10^{15}$ cm$^{-3}$, which is consistent with the idea that the depletion edge is in the tail of Zn diffusion.

The low reverse-bias leakage can be used to estimate the lifetime of carriers, which in the depleted diode region is typically limited by the Shockley-Read-Hall (SRH) generation-recombination processes associated with deep-level impurities. The SRH rate, $1/\tau_g$, can be estimated from the measured dark-current density due to SRH generation,

$$J = \frac{e n_i d}{\tau_g} \qquad (3)$$

The intrinsic carrier concentration in our InGaAsP diodes is about $7 \times 10^8$ cm$^{-3}$, based on the known $n_i = 1.3 \times 10^7$ cm$^{-3}$ of InP and the bandgap of our diodes, which is 100 meV narrower than that of InP. Using Eq. (3) with $J = 1$ nA/cm$^2$, we find $\tau_g \approx 20$ μs. This long SRH recombination time evidences high material quality in the epitaxial layers.

Next, we carried out optical testing of the photodiode. Figure 6 shows the transmission spectrum (*a*) of the entire structure comprising an InP substrate and the quaternary diode. Both absorption edges are clearly separated. On the same graph we plot transmission *luminescence* spectra of InP structure – with (*b*) and without (*c*) the epitaxial diode layers. Transmission luminescence is excited by red laser (640 nm) and measured from the opposite side of the wafer. The spectrum (*c*) is representative of the luminescence incident on the diode, when InP is excited far away from the diode interface. It is narrower than the intrinsic luminescence spectrum of InP because the short wavelength components of the latter are filtered out in the passage through the substrate. The spectrum (*b*) is excited in the diode and measured from the InP backside. For the rather excitation intensity used, the internal electric field of the *pin* diode is screened by the generated free carriers, so that one can regard spectrum (*b*) as representing the intrinsic luminescence of the quaternary material. It is largely undistorted in passage through the substrate since the diode material emits mostly in the region of InP transparency. The reflection luminescence spectrum of the diode (measured from the same side where it is excited by the red laser) is indistinguishable from its transmission luminescence.

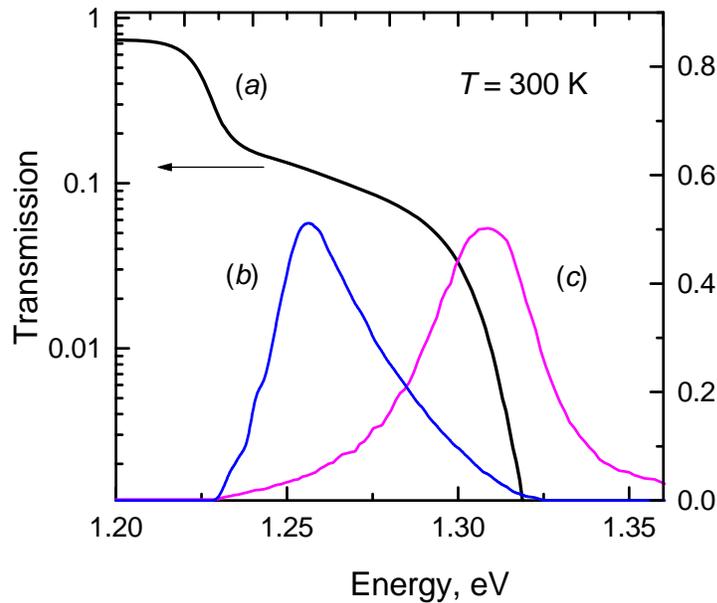

**Fig. 6**. Transmission spectrum (*a*) of the entire structure, comprising 350 μm thick InP substrate and 2 μm thick quaternary diode layers. Also shown (on the linear scale, normalized) are transmission luminescence spectra of the diode (*b*) and the bare InP substrate (*c*), both excited by a red laser (640 nm).

The luminescence spectrum (*c*) in Fig. 6 was taken in the bare InP structure without epitaxial layers; it is representative of the luminescence incident on the diode, when InP is excited far away from the diode interface. The spectrum is narrower than the intrinsic luminescence spectrum of InP because the short-wavelength components of the latter are filtered out in the passage through the substrate. In contrast, when we excite InP in the epitaxial structure, we observe "cascade" spectra, shown in Fig. 7. Red laser excites luminescence in InP substrate and then InP luminescence excites luminescence in the diode. Both the transmission and reflection luminescence spectra are shown in Fig. 7. In the transmission spectrum we see predominantly the luminescence of the diode material because diode layers effectively filter out the incident luminescence from InP. In contract, the reflection luminescence spectrum is dominated by the luminescence of InP substrate but also shows a peak in the red wing corresponding to the diode luminescence excited by InP transmission luminescence and propagating back through the substrate.

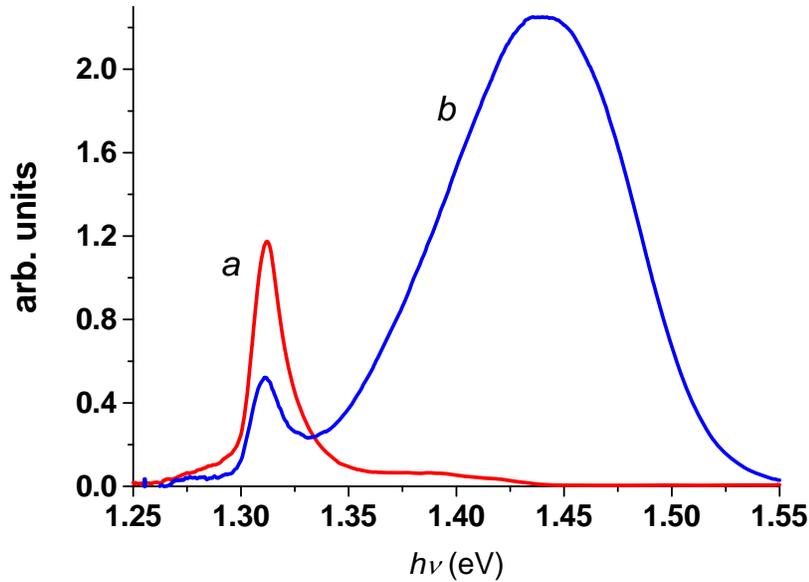

**Fig. 7**. Cascade luminescence spectra at 77 K, in the transmission (a) and reflection (b) geometries. Both spectra are excited by CW excitation at 640 nm, incident on the InP substrate. The InP luminescence excites luminescence in the diode. The reflection spectrum is dominated by InP luminescence but also contains the diode peak propagating back through the substrate.

Next, we present "optoelectronic" measurements in which the luminescence is excited in the InP substrate and the electrical response of the diode is measured. For characterization of the optical response of the fabricated photodiode, a front-end readout circuitry with charge-sensitive amplifier has been designed and directly interfaced to the photodiode to produce a voltage signal proportional to the total charge created in photodiode. We performed two sets of experiments in an effort to fully characterize designed photodiode, first ones in which the incident photons were generated by the laser and the second set of experiments with the real radiation sources. Radiation tests will be described in the next section.

Laser excitation is useful because we can control and vary the number of photons incident to the sensor. An additional advantage of laser experimentation is that the laser beam can be split and the number of photons in individual pulses can be simultaneously measured with a calibrated commercial silicon detector. Another attractive feature of the laser experiments is that with changing wavelength, we can change the penetration depth of the photons in the detector, i.e. the location where the luminescence photons are created in InP. The side of the sensor that was irradiated was the back side of the detector, side opposite to the epitaxial diode. At the wavelength of 970 nm (1.28 eV), bulk InP material becomes transparent and the electrons in photodiode are generated directly by the laser photons.

To check the linearity of the photodiode response, we made a set of measurements where the electrons in photodiode were created by 1000 individual pulse signals at the same nominal energy of the laser. The amount of photons incident on the sensor in each set of measurements was varied by inserting attenuation filters of different magnitudes between the laser and the sensor. The energy of individual pulses varies about 10% of the nominal value due to laser jitter. The number of incident laser photons takes into account Fresnel reflection and was estimated based on the power measured by silicon detector in the exact location of our detector, without attenuation filters in the path of laser beam. In Fig. 8 the mean measured number of electrons is plotted as a function of the number of incident photons. This curve relates the number of electrons recorded to the number of incident photons and represents the efficiency of photoelectric conversion. Our setup was not optically tight and at very low incident energy the linearity is disrupted by scattered background light.

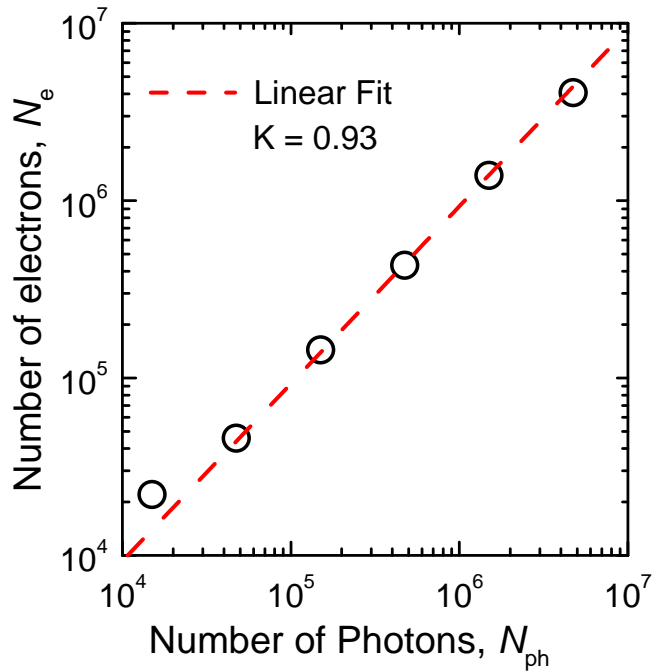

**Fig. 8**. Mean number of electrons $N_e$ registered by the detector at room temperature as a function of incident energy on the detector in terms of the number of incident photons $N_{ph}$ at wavelength of 970nm. The dashed line represents a linear fit with $K = dN_e/dN_{ph}$.

## 4. Preliminary ionizing radiation experiments

The intended application of the proposed epitaxial photodiode is as optically-tight photoreceiver on the surface of semiconductor scintillator. To examine the performance of fabricated photodiode as scintillator's photoreceiver, the resolution of radiation event quantification is estimated. This is achieved by estimation of the theoretical noise floor in detection of an ionizing radiation event.

Photodiode characteristics that are essential in determination of theoretical noise floor are the associated parallel capacitance $C_{det}$, for the fabricated diode fairly large and measured at 55 pF, and leakage current $I_{LK}$, measured at 10 pA. For a single radiation event in scintillator, signal produced by photodiode can be modeled as a short charge pulse. To convert this charge signal into voltage signal, a readout circuitry has to be interfaced with the photodiode. We adopt the readout system [15,16] that consists of a charge sensitive amplifier (CSA) that performs charge-to-voltage conversion and a pulse shaper, implemented as a bandpass filter in order to achieve the optimal signal-to-noise ratio at the output of the system. Calculation of the noise floor is based on the assumption that the dominant noise sources are the shot noise of the photodiode and noise (thermal and flicker noise) of the input MOS transistor of CSA directly interfacing the photodiode. The input MOS transistor dominates the noise introduced by the readout electronics due to subsequent signal amplification and hence its optimization leads to optimal sensitivity. Noise performance of the system is quantified through equivalent noise charge (ENC). The ENC is defined as the ratio of the total integrated rms noise at the output to the signal amplitude due to one electron charge and is directly proportional to the minimum detectable signal. To register the radiation event, pulses produced as response to the event by the readout circuitry, have to be detected using simple threshold function. The threshold determines minimum detectable charge and is set to value between 3 and 7 multiples of ENC, depending on desired ratio of false positives and false negatives. Equivalent noise charge (ENC) can be calculated [15,16] as

$$ENC^2 = (C_{det} + C_{OX}WL)^2 \left[ \frac{a_{th}}{\tau} 4kT \cdot \frac{2}{3g_m} + a_f \frac{K_f}{WLCox \cdot f} \right] + a_{sh}\tau 2qI_{LK}, \qquad (4)$$

where $C_{OX}$ is the oxide capacitance per unit area, $W$, $L$ and $g_m$ are width, length and transconductance of input transistor and $K_f$ is the *1/f* noise coefficient. The constants $a_{th}$, $a_f$ and $a_{sh}$ depend on the order of filtering performed in the optimization of signal-to-noise ratio in the readout system, The time constant $\tau$ of the filtering inversely scales the thermal noise contribution of input transistor, while directly scales the shot noise contribution of the diode. Additional constraints in the optimization of the input transistor for specified leakage current and capacitance of diode are the power consumption, chip area and the rate of the radiation events, which sets the upper limit for the choice of time constant $\tau$. The obtained theoretical minimal value of ENC, based on the technology parameters for 0.5μm CMOS process chosen for readout circuit implementation, is 97 electrons, with the optimal time constant equal to 24 μs. However, to obtain this value of ENC, the required biasing current of the input transistor is 29 mA, which would lead to prohibitive power consumption, especially in the proposed 3D scintillator array. For more acceptable biasing currents, the value of ENC is increased, e.g. for the biasing current of 1 mA, the ENC is 112 electrons.

Due to the large parasitic capacitance, the thermal noise component dominates the equivalent noise in (4). The main emphasis in the design of the photodiode was on the reduction of the leakage current. By increasing the value of time constant $\tau$, the thermal noise contribution decreases and if the leakage current is small, there is no significant penalty in increased shot noise with increased $\tau$. The dependence of ENC on the diode's leakage current is illustrated in Fig. 9 for a fixed biasing current of the input transistor at value of 1 mA. The figure demonstrates the significance of minimizing leakage current for high-resolution quantification of radiation events. Note, however, that at very low leakage currents, the ENC increase only moderately with the leakage, since in that range the dominant noise component is thermal noise, proportional to the diode capacitance.

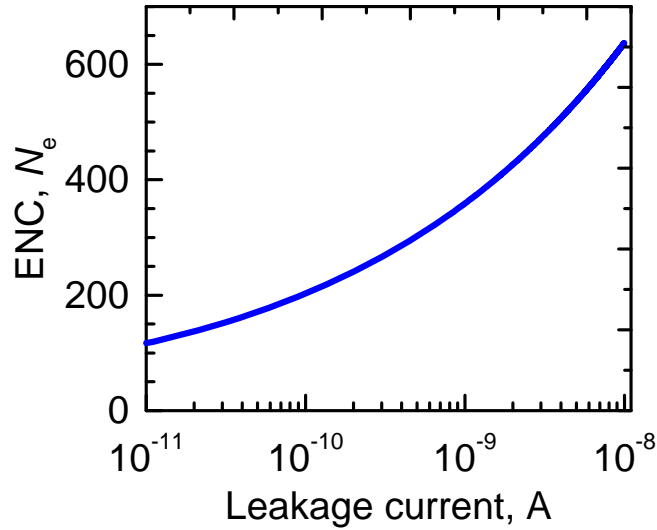

**Fig. 9**. Dependence of the equivalent noise charge (ENC) on the leakage current of the photodiode calculated for a fixed diode capacitance of 55 pF and fixed biasing current (1 mA) of the input MOS transistor of preamplifier. The ENC is a crucial noise characteristic. The minimum detectable signal is represented in multiples of ENC, with the multiplicative value determined by the desired balance of false positives and true negatives in the discrimination of radiation events.

The response of the fabricated photodiode to gamma radiation was characterized with a CSA directly interfaced to the photodiode. The output signal of CSA is a voltage signal proportional to the total charge created in photodiode; it is recorded using NI-6259 data acquisition card from National Instruments. We performed experiments with several radiation sources.

***Radiation experiments with $^{241}$Am.*** We performed a set of experiments with americium $^{241}$Am (a real radiation source, found in smoke detectors). Americium was placed at a distance of 2 mm from the fabricated detector in a metal enclosure with measurement system. Americium emits high-energy alpha particles (5 MeV), and low-energy gamma particles (60 keV and lower). Therefore, in the experiments, the response of the detector corresponds to the high-energy alpha particles. Longer recordings were performed to obtain statistical information on the energy of emitted alpha particles. Figure 10 shows the recorded distribution of energy of alpha particles as a histogram. The *x*-axis is the number of electrons, and the *y*-axis is the percentage of the total recorded population in each bin. Figure 10 (*a*) shows the recorded energy distribution when the source was facing the back side of the detector, opposite to the epitaxial diode. Figure 10 (*b*) shows the recorded energy distribution when the source was placed facing the top side of the detector, the side with epitaxial photodiode. The recording time was 200 s. Total number of detected particles was 3741 at 77 K and 919 at 300 K, when back side is facing the radiation source, and 26,523 at 77 K and 5194 at 300 K, when front side is facing the source. Statistical information on distribution of energy of α-particles at 77K differs from that obtained at 300K. From the figure, we observe the shift of the distribution at low temperature toward higher energy, as expected due to higher scintillator yield at lower temperatures.  Penetration depth of alpha particles is around 10 μm, so in the first case, electrons are created on the surface of the InP bulk opposite to the photodiode, while in the second case the electrons are in photodiode and layer of InP adjacent to the photodiode. The expected number of electrons created with alpha particles is around $10^6$, since there is additional attenuation of alpha particles due to packaging of the source (thin layer of metal).

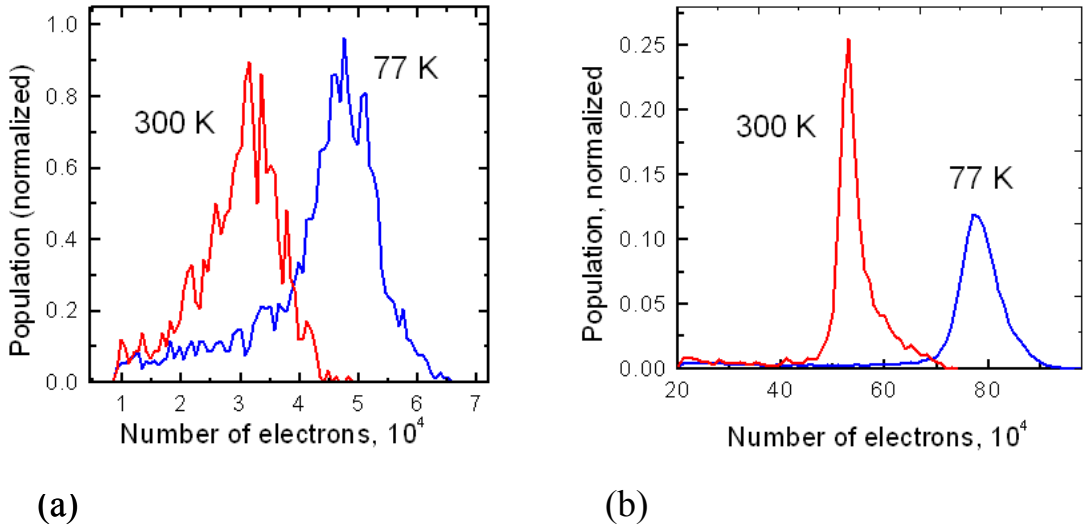

(a)                                                       (b)

**Fig. 10**.  Histogram of alpha particle energy distribution at 77 K and 300 K when radiation source is facing (*a*) back and (*b*) top of the detector.

*Radiation experiments with* $^{137}$**Cs.** To demonstrate the scintillator response to γ-quanta, we have also performed radiation experiments with the radioactive isotope $^{137}$Cs that produces γ quanta of energy 662 keV. The experiments were performed at room temperature and temporal waveforms of the response to single γ-quanta are shown in Fig. 11. The response is exactly as estimated: a step response followed by an exponential decay with time constant of 140 μs, which is the time constant of resistor and capacitor in the feedback loop of high-gain amplifier in the CSA.

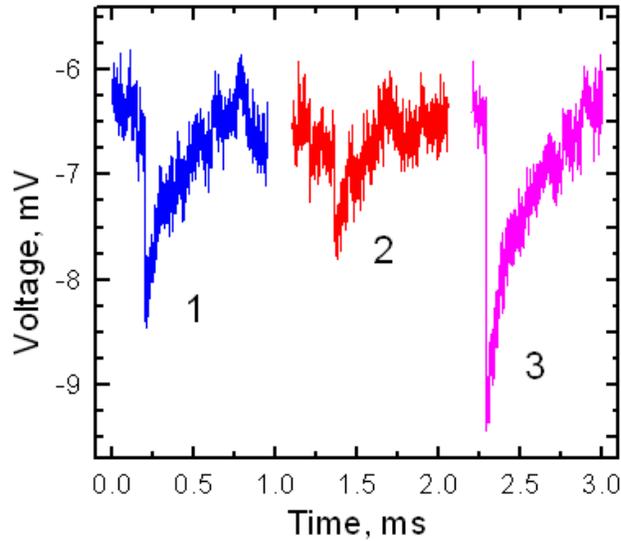

**Fig. 11**. Temporal response of the photodiode when the scintillator is excited by $^{137}$Cs isotope. Three traces shown correspond to distinct Compton events. The effective capacitance of the charge-to-voltage circuit is 1.4 pF and the three successive events correspond to charges of 11, 6.5, and 24 $\times 10^3$ electrons, respectively.

## 5. Conclusion

We report the design, fabrication and operation of quaternary InGaAsP photodiode, epitaxial on InP and sensitive to InP luminescence. Owing to the achieved small leakage current of 10 pA, high-resolution quantification of radiation events is feasible with the designed photodiode, despite its large parallel capacitance – with value of the noise floor on the level of a hundred of electrons. We have demonstrated registration of single alpha particles and γ-quanta in a InP scintillator with fabricated photodiode on the surface of the scintillator.

## Acknowledgments

This work was supported by the Domestic Nuclear Detection Office (DNDO) of the Department of Homeland Security, by the Defense Threat Reduction Agency (DTRA) through its basic research program, and by the New York State Office of Science, Technology and Academic Research (NYSTAR) through the Center for Advanced Sensor Technology (Sensor CAT) at Stony Brook.